\documentclass[aip, pof, preprint]{revtex4-1}
\usepackage{graphicx}
\usepackage{color}
\usepackage{amssymb}
\usepackage{amsmath}
\usepackage{tabularx}
\usepackage{bm}
\usepackage{hyperref}
\usepackage{ltxtable}
\usepackage{natbib} 
\usepackage{longtable}
\newcommand{\be}{\begin{equation}}
\newcommand{\ee}{\end{equation}} 
\newcommand{\bea}{\begin{eqnarray}}
\newcommand{\eea}{\end{eqnarray}}

\usepackage[usenames,dvipsnames]{xcolor}
\hypersetup{
colorlinks,
citecolor=blue,
filecolor=blue,
linkcolor=blue,
urlcolor=blue}

\begin{document}
\title{Large eddy simulations using renormalized viscosity}

\author{Sumit Vashishtha}
\email{sumitv@iitk.ac.in}
\affiliation{Department of Physics, Indian Institute of Technology, Kanpur 208016, India}
\author{Anando Gopal Chatterjee}
\email{anandogc@iitk.ac.in}
\affiliation{Department of Physics, Indian Institute of Technology, Kanpur 208016, India}
\author{Abhishek Kumar}
\email{ac7600@coventry.co.uk}
\affiliation{Applied Mathematics Research Centre, Coventry University, Coventry CV15FB, The United Kingdom}
\author{Mahendra K. Verma}
\email{mkv@iitk.ac.in}
\affiliation{Department of Physics, Indian Institute of Technology, Kanpur 208016, India}
\date{\today}

\begin{abstract}
In this paper, we have used renormalized viscosity for performing large eddy simulations (LES) of decaying homogeneous and isotropic turbulence inside a periodic cube on coarse grids ($32^3$, $64^3$ and $128^3$) at initial Taylor Reynolds number ${\mathrm{Re}_{\lambda} \approx 315} $. The LES results were compared with direct numerical simulation (DNS) results on $512^3$ grid.  We observe good agreement between  LES and DNS results on the temporal evolution of turbulence kinetic energy ${E(t)}$, kinetic energy spectrum ${E_{u}(k)}$, kinetic energy flux ${\Pi_u(k)}$.  The isosurfaces of velocity magnitude at a given time are similar for DNS and LES. This establishes the suitability of using renormalized viscosity for performing large eddy simulations.

\end{abstract}

% 47.27.te Turbulent convective heat transfer
% 47.55.P- Buoyancy-driven flows; convection

\maketitle
\section{Introduction}
\label{sec:intro}
Turbulence remains to be one of the most challenging problems in classical physics. A direct numerical simulation (DNS) at  extreme Reynolds numbers ($\mathrm{Re}$) remains infeasible even on some of the fastest contemporary supercomputers. This is because of a rapid increase in the range of scales (${N\propto \mathrm{Re}^{9/4}}$) that must be resolved in DNS for large $\mathrm{Re}$. Large eddy simulation (LES) is one of the most efficient techniques for simulating turbulent flows. 
 
 In LES, only the large scales of turbulent flows are simulated, and the unresolved scales are appropriately modeled~\cite{doi:10.1146/annurev.fluid.32.1.1}.  In turbulence, Fourier modes corresponding to different length scales interact with each other. According to Kolmogorov's~\cite{Kolmogorov:DANS1941Dissipation, Kolmogorov:DANS1941Structure} theory of turbulence, the nonlinear interactions in a turbulent flow yield a constant energy flux from large scales to intermediate scales, and then to small scales. When we observe a fluid flow at length scales greater than $l$, where $l$ belongs to inertial range, then the the energy flux $\Pi$ equals the energy dissipation rate.  Also, the effective viscosity at length scale $l$ is proportional to $\Pi^{1/3} l^{4/3}$.  Physically, the momentum diffuses with above enhanced viscosity.  This idea is exploited in LES.

 The earliest SGS model used was by Smagorinsky~\cite{smagorinsky1963general} who modelled the effect of small scales  using an eddy viscosity:
 \be
 \nu_\mathrm{Smag} = (C_s \Delta)^2 \sqrt{2 \bar{S}_{ij} \bar{S}_{ij}}
 \label{eq:Smag}
 \ee
 where $\bar{S}_{ij}$ is the stress tensor of the resolved scales, $\Delta$ is the smallest grid scale, and $C_s$ is a constant that is taken between 0.1 and 0.2.   Certain issues with Smagorinsky model such as unconditional dissipation, neglect of backscatter, and empirical nature of the constants involved are addressed in the dynamic Smagorinsky model~\cite{germano1991dynamic} in which the subgrid scale stresses at two different filtered levels and the resolved turbulent stresses are utilized to evaluate the effective viscosity.  Another class of LES models exploit  scale-similarity~\cite{sarghini1999scale} of the flow structures above and below the cutoff. This scheme is  exploited to distinguish subgrid scales from supergrid scales. Several other LES models~\cite{domaradzki1997subgrid,scotti1999fractal,Misra:PF1997} focus on  SGS velocity field rather than SGS tensor. In the works of Misra and Pulin~\cite{Misra:PF1997} and Cheng et al.~\cite{Cheng:JFM2015},  the SGS structure of the turbulence is assumed to be the stretched vortices whose orientations are determined by the resolved velocity field. The implied velocity field is used to evaluate the SGS stress tensor.  There are variations of the above models, as well as attempts to fine-tune the models for complex flows involving confined and complex geometries, boundary layers, etc.
 
 A less popular LES model is based on the renormalised viscosity computed using renormalisation group (RG) analysis~\cite{Yakhot:JSC1986,McComb:book:Turbulence,McComb:book:HIT}.  RG helps us understand problems with multiple scales, turbulence being one such problem.   In Wilson's Fourier-space RG scheme, Fourier space is divided into many shells, and the nonlinear interactions among various shells are computed using first-order perturbation theory that yields scale-dependent viscosity, called renormalized viscosity.   Here we state the formula for the renormalized viscosity $\nu_\mathrm{ren}$ derived by McComb~\cite{McComb:book:Turbulence,McComb:book:HIT}
  \be
\nu_\mathrm{ren}(k) = K_\mathrm{Ko}^{1/2} \Pi^{1/3} k^{-4/3}\nu_*,
\label{eq:nu_r0}
\ee
where $k$ is the wavenumber, $K_\mathrm{Ko}$  is the Kolmogorov's constant, and $\nu_*$ is a constant. 
Using RG computation, McComb and Watt~\cite{McComb:PRA1992Two_field} found that $\nu_* \approx 0.50$ and $K_\mathrm{Ko} \approx 1.62$,  Verma~\cite{Verma:PR2004} also computed the above quantities using a refined technique and found $\nu_* \approx 0.38$ and $K_\mathrm{Ko} \approx 1.6$. Note that the above formula has been derived from the first principles (from the Navier-Stokes equation) for homogeneous and isotropic turbulence under certain assumptions~~\cite{McComb:book:Turbulence,McComb:book:HIT}.
 
 This renormalized viscosity of Eq.~(\ref{eq:nu_r0}) is very similar to that employed in Smagorinsky model (see Eq.~(\ref{eq:Smag})).  For a subgrid cutoff of $\Delta$ (in real space), the wavenumber cutoff is $k_c=\pi/\Delta$.  Hence, the renormalized viscosity to be employed for LES would be
   \be
 \nu_\mathrm{ren}(k) = K_\mathrm{Ko}^{1/2} \Pi^{1/3}k_c^{-4/3}\nu_* .
\label{eq:nu_r}
\ee
 We can easily demonstrate equivalence between Eq.~(\ref{eq:Smag}) and Eq.~(\ref{eq:nu_r}).   Equation~(\ref{eq:Smag}) is converted to Fourier space as
 \bea
 \nu_\mathrm{Smag} & = & \left( C_s \frac{\pi}{k_c}\right)^2 
 \left[ \int_0^{k_c} 2 k^2 E(k) dk \right]^{1/2} \nonumber \\
 & = &  \left( C_s \frac{\pi}{k_c}\right)^2  \left[ \frac{3}{2} K_\mathrm{Ko} \Pi^{2/3} k_c^{4/3} \right]^{1/2} \nonumber \\
 & = &  C_s^2 \pi^2  \left[ \frac{3}{2} K_\mathrm{Ko} \right]^{1/2} \Pi^{1/3} k_c^{-4/3}, 
 \eea
 where $E(k)$ is the energy spectrum, which is taken to be Kolmgorov's spectrum as an approximation.  Now a comparison of Eq.~(\ref{eq:Smag}) and Eq.~(\ref{eq:nu_r}) yields
 \be
 C_s = \frac{\nu_*^{1/2}}{\pi}  \left( \frac{2}{3}\right)^{1/4}.
 \ee
 For $\nu_* \approx 0.38$, we obtain $C_s \approx 0.177$, which lies in the range of values employed in LES.   The above computation shows usefulness of RG scheme to compute the undetermined constants of LES.  
 
Most of the RG computations however are for homogeneous and isotropic turbulence~\cite{carati1989renormalization}, hence the constants (e.g $\nu_*$,$K_\mathrm{Ko}$) computed using these calculations are not suitable for anisotropic and inhomogeneous flows. Hence, an extensive work is required for realistic estimates of the parameters from the first-principle calculations.  As a first step, it is important to validate the subgrid viscosity model of  Eq.~(\ref{eq:nu_r}) using numerical simulations, that is to compare the results of  DNS and LES. Verma and Kumar~\cite{Verma:Pramana2004}  performed one such analysis for decaying hydrodynamic turbulence in a periodic box.  They showed that the  evolution of total energy, as well as the energy spectrum, in DNS of $128^3$ and LES of $64^3$ are in good agreement with each other.  In this paper we compare results of DNS on finer grids ($512^3$) with LES results on coarser-grids ($32^3$, $64^3$, $128^3$).  In the present work we have performed extensive validation tests for LES. For example, we show that in addition to energy spectrum and energy evolution, the energy fluxes of LES and DNS match quite well.  We also show bottleneck effect in LES. Thus, the LES scheme presented in this paper is more refined than that of Verma and Kumar~\cite{Verma:Pramana2004}.

The outline of the paper is as follows: In Sec.~\ref{sec:Maths}  we discuss the details regarding the governing equations and the renormalized viscosity used in LES. Computational methodologies are discussed in Sec.~\ref{sec:CM}. The results obtained from LES and DNS are discussed in Sec.~\ref{sec:Results}. Finally, we summarize our results in Sec.~\ref{sec:Conclusion}.
     
%%%%     
\section{LES formulations using renormalized parameters}
\label{sec:Maths}

In this section, we present the formalism of LES using renormalized parameters.  The incompressible Navier-Stokes equation in real space is given by
\bea
\label{eq:NSEreal}
 \frac{\partial \mathbf{u}}{\partial t} + \mathbf{u}.\nabla\mathbf{u} & =& -\nabla p + \nu_0{\nabla^2}\mathbf{u}\\
 \label{eq:divergence_real}
 \nabla.\mathbf{u}&=&0
\eea
where $\mathbf{u}$ is velocity vector, p is the pressure and $\nu_0$ is the kinematic viscosity.  However, decimation of smaller scales is more convenient in Fourier space, hence most of the  RG analysis work on turbulence have been performed in the Fourier space. We write the corresponding incompressible fluid-flow equations in Fourier space as
\bea
\label{eq:Fourier_NSE}
(\frac{d}{dt} + \nu_0 k^2) \hat{u}_j(\mathbf{k},t) & = & -\frac{i }{2} k_j P_{jkl}(\mathbf{k})\sum_{\mathbf{k',k''}}\delta_{\mathbf{k,k'+k''}}\hat{u}_k(\mathbf{k'})\hat{u}_l(\mathbf{k''}) \\
\label{eq:divergence_Fourier}
k_i u_i ({\bf k}) & = & 0,
\eea
where
\bea
P_{jkl}(\mathbf{k}) &=& k_kP_{jl}(\mathbf{k}) + k_lP_{jk}(\mathbf{k})\\
P_{jk}(\mathbf{k})&=& \delta_{jk} - \frac{k_j k_k}{k^2}
\eea
The right hand side of Eq.~(\ref{eq:Fourier_NSE}) represents the triadic interactions among the wavenumbers $\mathbf{k', k'', k}$ that satisfies $\mathbf{k' + k''= k}$.  Note the definition of the Fourier transform is
\be
{\mathbf{u}(\mathbf{x},t)= \sum_{\mathbf{k}}\mathbf{\hat{u}}(\mathbf{k},t)e^{i\mathbf{k.x}}}
\label{eq:ufourier}
\ee

In RG scheme, the Fourier modes of wavenumber shells are truncated in steps~\cite{Yakhot:JSC1986, McComb:book:RG,Zhou:PRA1988} that leads to elimination of some of the triadic interactions.  RG computation takes into account these interactions, and put these effects into an enhanced viscosity. It has been shown that the total effective viscosity at wavenumber $k$ is~\cite{McComb:book:RG}
   \be
 \nu(k)  = \nu_0 +  \nu_\mathrm{ren}(k) =\nu_0 + K_\mathrm{Ko}^{1/2} \Pi^{1/3}k_c^{-4/3}\nu_*.
\label{eq:nu_total_RG}
\ee
For details of RG procedure, refer to~\cite{Yakhot:JSC1986,McComb:book:RG,Zhou:PRA1988}.

The LES scheme based on renormalized viscosity makes use of Eq.~(\ref{eq:nu_total_RG}).  We employ a sharp spectral filter at cutoff wavenumber ${k_c}$:
\be
{\mathbf{\hat{\bar{u}}}(\mathbf{k}, t)=H(k_c -k)\mathbf{\hat u}(\mathbf{k}, t),}
\ee
where ${H}$ represents Heaviside function, and ${k}=|\mathbf{k}|$ is the magnitude of wavenumber.  Hence, the real space velocity is
\be
{\mathbf{\bar{u}}(\mathbf{x},t) = \sum_{\mathbf{k}}e^{i\mathbf{k.x}}\mathbf{\hat{\bar{u}}}(\mathbf{k}, t)=\sum_{|\mathbf{k}|<|\mathbf{k_c}|}e^{i\mathbf{k.x}}\mathbf{\hat{{u}}}(\mathbf{k}, t) }
\ee
With this, the equations for the resolved Fourier modes are 
\be
{(\frac{d}{dt} + \nu_\mathrm{tot} k^2)\hat{\bar{u}}_j(\mathbf{k},t)= -\frac{i}{2} k_j P_{jkl}(\mathbf{k})\sum_{\mathbf{|k'|,|k''|<|k_c|}} \delta_{\mathbf{k,k'+k''}}{H(k_c -k)}\hat u_k(\mathbf{k'}, t)\hat u_l(\mathbf{k''}, t)} .
\label{eq:cutoffeq}
\ee
Here, the effects of truncated modes ${\bf u(k)}$ with $k > k_c$ is accounted for by using renormalized viscosity~\cite{Yakhot:JSC1986, McComb:book:RG,Zhou:PRA1988,Verma2005:review}  is 
\be
\nu_\mathrm{tot} = \nu_0 +  \nu_\mathrm{ren}(k) =\nu_0 + K_\mathrm{Ko}^{1/2} \Pi^{1/3}k_c^{-4/3}\nu_*,
\label{eq:nu_LES}
\ee
where $\nu_* = 0.38$, $K_\mathrm{Ko}$ is the Kolmogorov constant, and $\Pi$ is the kinetic energy flux in the inertial range of wavenumbers. Note that renormalized viscosity is $\nu_\mathrm{ren}(k) $ computed at the cutoff wavenumber $k_c$, and that the $k_c$ is assumed to lie in the inertial range.

Now several important issues regarding LES implementation is in order.  The computation of $\nu_\mathrm{tot}$ for LES requires the Kolmgorov's flux $\Pi(k_0)$, where $k_0$ is in the inertial range.  In our simulation, we compute $\Pi(k_0)$ using the mode-to-mode formula of Verma~\cite{Verma:PR2004} and Dar et al.~\cite{Dar:PD2001}:
 \be
 {\Pi_u{(k_0)} = \sum_{k\geq k_0}\sum_{p< k_0}\delta_{k,p+q}\mathrm{Im}[\mathbf{k.u(q)}][\mathbf{u^{*}(k).u(p)}]}.
 \label{eq:Pi}
\ee
Regarding the choice of $k_c$ in a $N^3$ periodic box simulation, we take $k_c=N/3$ due to dealising employed in our DNS and LES. Under the 2/3 rule of dealising, the Fourier modes $|{\bf k}| > N/3$ are set to zero (Note that $k_\mathrm{max} = N/2$).  Hence, the nonzero Fourier modes are with $k_i = [-N/3:N/3]$, where $i=x,y,z$.  Therefore, for our DNS and LES, the effective $k_\mathrm{max} = N/3$, not $N/2$. We employ the above $k_c = N/3$ for our LES simulations.  We remark that these schemes are superior to those employed by Verma and Kumar~\cite{Verma:Pramana2004} who employed $k_c=N/2$, and the viscous dissipation rate as  an estimate for the energy flux $\Pi$.

In Sec~\ref{sec:CM} we discuss different computational methodologies associated with our simulations.

\section{Simulation details}
\label{sec:CM}
 We solve Eqs.~(\ref{eq:NSEreal},~\ref{eq:divergence_real}) in our DNS and LES.  We employ pseudo-spectral method for our simulation and use the code Tarang~\cite{Verma:Pramana2013tarang}.  We perform DNS computations on $512^3$ grid, and LES computations on $32^3$, $64^3$ and $128^3$ grids.  For these simulations, we use a periodic cube of size $2\pi\times2\pi\times2\pi$, hence the wavenumber components are integers.  In our simulations, time-marching is done using fourth-order Runge-Kutta method. Furthermore, the $2/3$ rule~\cite{Canuto:book:SpectralFluid} is used for dealiasing, and the Courant-Friedrichs-Lewy (CFL) condition is employed for determining the time step ${\Delta t}$.  In all our simulations, we take the kinematic viscosity $\nu_0=10^{-3}$.
 
 First we perform a forced DNS on $512^3$ grid with $\nu_0=10^{-3}$.  We let the flow evolve to a steady state.  We use the steady-state flow profile as an initial condition for DNS on $512^3$ grid, as well as for the LES simulations on coarser grid. Under this scheme, the Fourier modes of the LES (at the resolved scales) are exactly same as those in DNS.   The Reynolds number based on Taylor's micro scale, $R_{\lambda} \approx 315$ for the initial condition.  Starting from these initial conditions, we perform decaying DNS   and  LES simulations.  For the LES, we take $k_c=N/3$ and viscosity as given in Eq.~(\ref{eq:nu_LES}).  Of course, for the DNS, $\nu_\mathrm{tot} = \nu_0$.  The decaying simulations have been carried out till non-dimensional time $t=50$; here the time unit is $L/U$, where $L,U$ are the large scale length and velocity.
 
In the following section we  compare the results of DNS and LES.

%%%%%
\section{Comparison of DNS and LES results}
\label{sec:Results}

In this section, we compare the DNS and LES results on the evolution of global quantities such as total energy and total dissipation rate, the energy spectrum and flux, as well as real space profile. First we start with the evolution of total energy $E(t)$ and dissipation rate $\epsilon(t)$, which are defined as
\bea
\label{eq:E(t)}
 E(t) &=& \frac{1}{2}\sum_{k} |\bar{\bf u}(\mathbf{k},t)|^2\\
 \label{eq:dissipation}
\epsilon(t) & = &  \sum_{k} 2\nu_{\mathrm{tot}} k^2 E_u(k)=\sum_{k}2(\nu_0 + \nu_{\mathrm{\mathrm{ren}}}(k_c))k^{2}E_u(k),
\eea
 where $\bar{\bf u}(\mathbf{k},t)$ represents the Fourier components of the resolved velocity. For DNS, $\bar{\bf u}(\mathbf{k},t)$ is the full velocity field.  
 
\begin{figure}
\begin{center}
\includegraphics[scale=0.4]{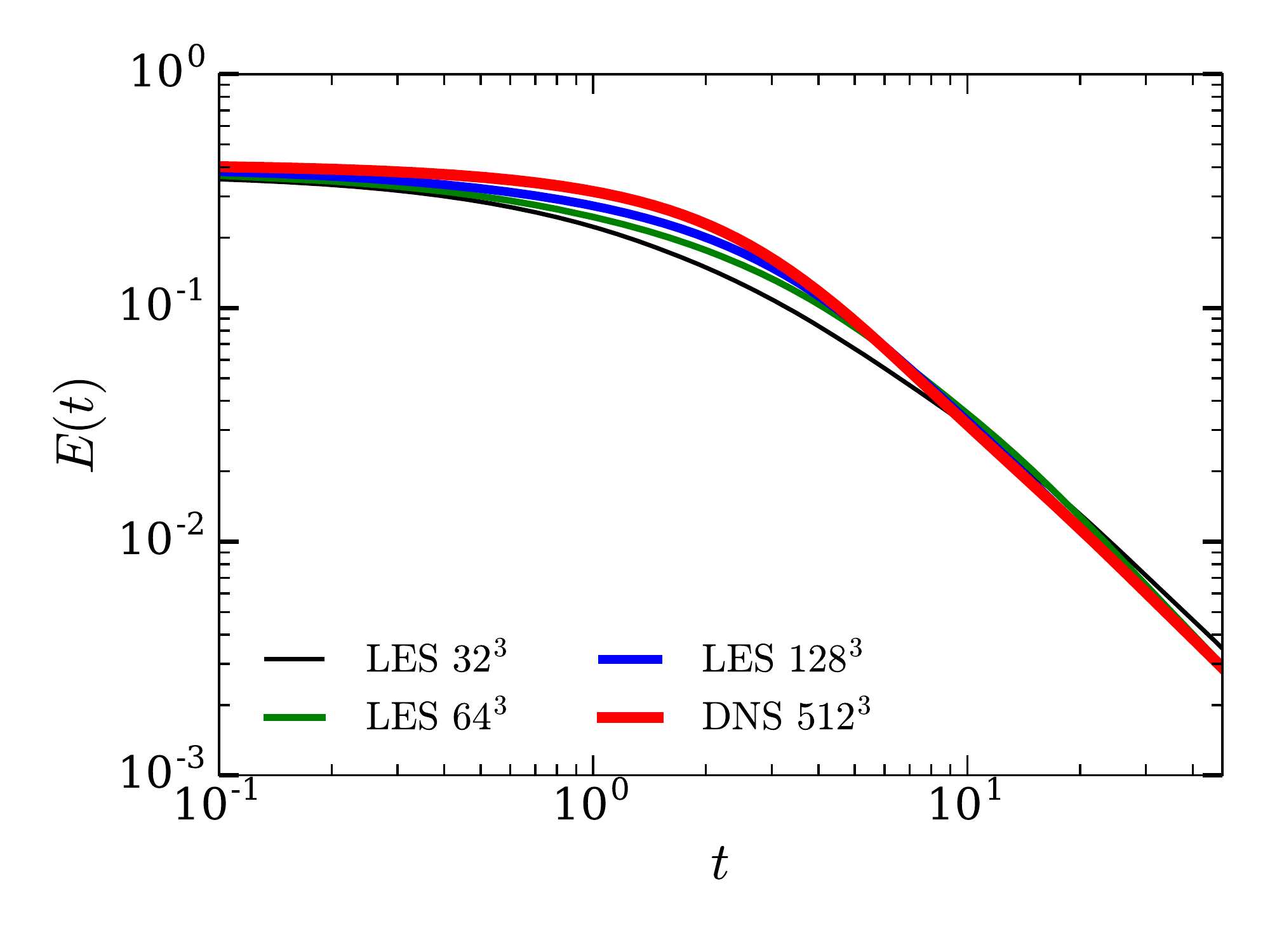}
\end{center}
\caption{Temporal evolution of turbulence kinetic energy $E(t) = u^2/2$. Note that the evolution of $E(t)$ is similar for LES and DNS runs.}
\label{fig:KE} 
\end{figure}
\begin{figure}
\begin{center}
\includegraphics[scale=0.4]{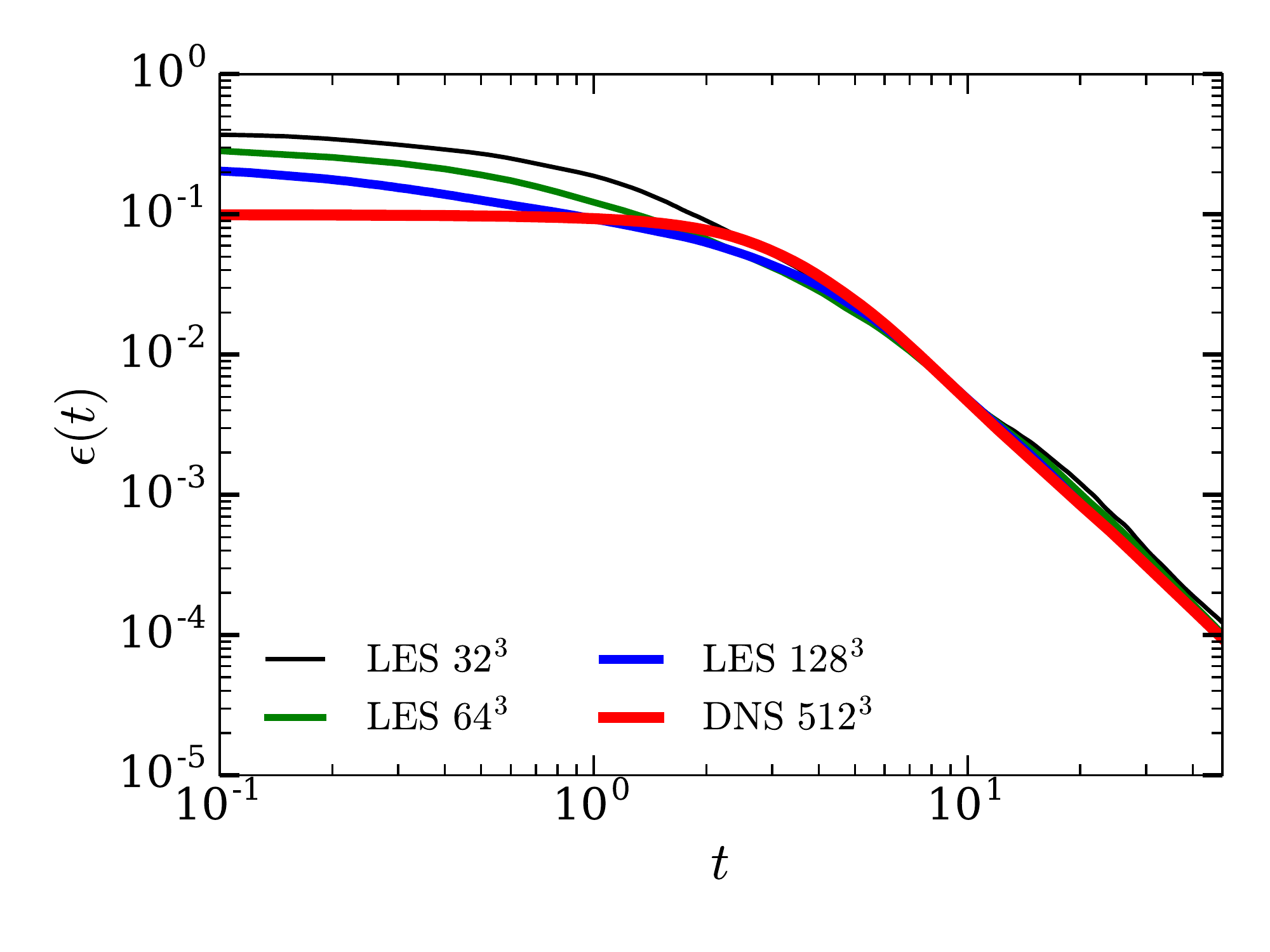}
\end{center}
\caption{Temporal evolution of total viscosity ${\nu_{\mathrm{tot}} = \nu_0 + \nu_{\mathrm{ren}}}$. For all the runs, $\nu_{\mathrm{tot}} $ decreases with time, and  it reaches $\nu_0$ asymptotically.  $\nu_{\mathrm{tot}} $ is largest for LES $64^3$ due to Eq.~(\ref{eq:nu_LES}).}
\label{fig:nu} 
\end{figure}
In Fig.~\ref{fig:KE}, we exhibit the temporal evolution of total turbulent kinetic energy  ${E(t)}$ for DNS and LES. We observe that $E(t)$ for all the runs are  very similar.   Note however that the initial energy of a lower-resolution LES  is smaller than that of DNS and higher-resolution LES.  This is because of the fewer number of modes present in the lower-resolution runs.  In Fig.~\ref{fig:nu} we illustrate the total  viscosity ${\nu_{\mathrm{tot}} = \nu_0 + \nu_{\mathrm{ren}}}$ as a function of time.  Clearly, $\nu_{\mathrm{tot}}$ is largest for the LES lowest grid ($64^3$) because $k_c$ is smallest for this run.  Also, $ \nu_{\mathrm{tot}}= \nu_0$ for DNS.  As time progresses, the total energy decreases for all the runs, and the flow becomes viscous.  At $t=50$, $\mathrm{Re_{\lambda}} \approx 48$  for all the flows.    Therefore,  asymptotically,  $ \nu_{\mathrm{tot}} \rightarrow \nu_0$.

\begin{figure}
\begin{center}
\includegraphics[scale=0.4]{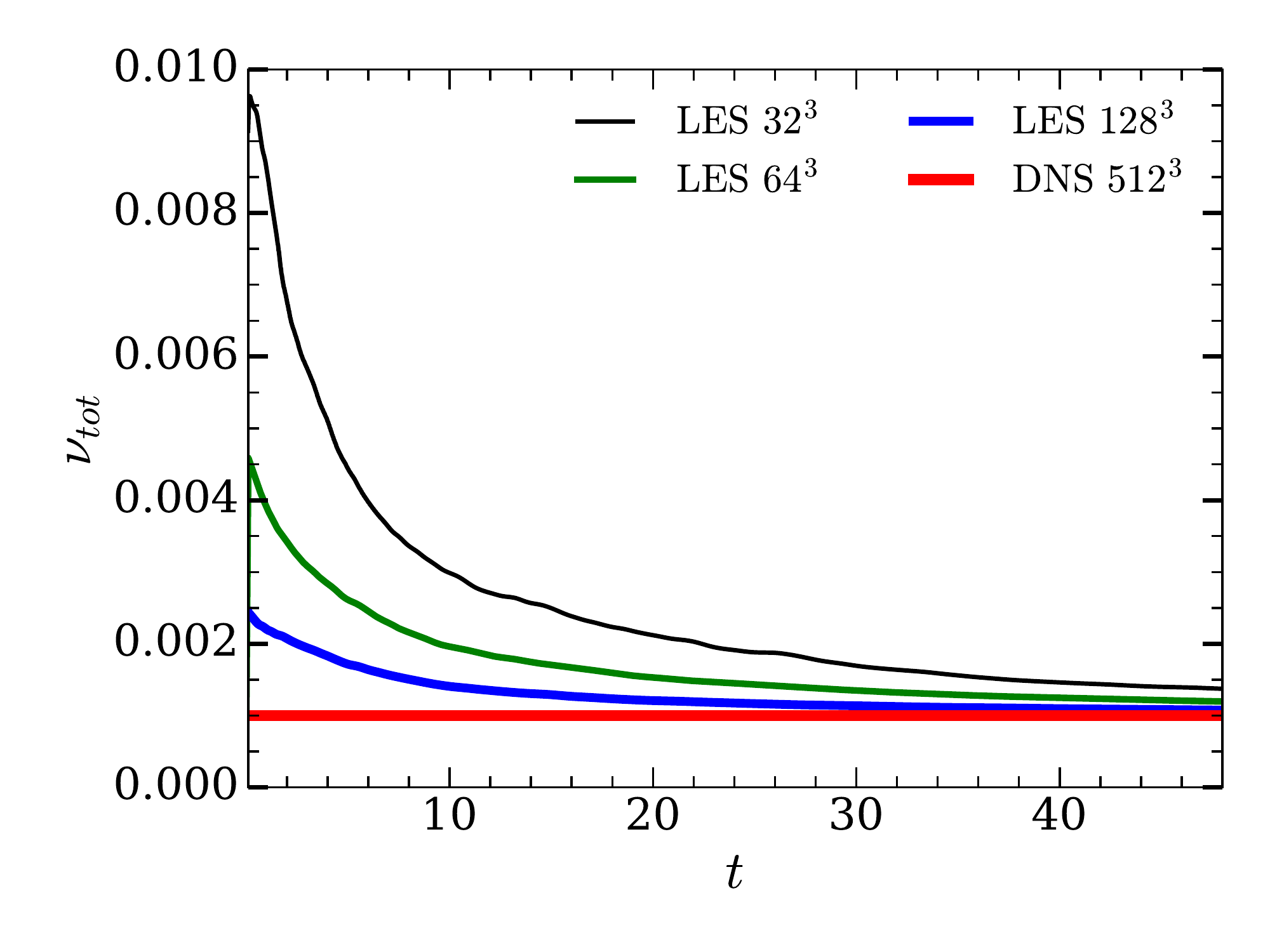}
\end{center}
\caption{Temporal evolution of Total dissipation $\epsilon(t)$. The $\epsilon(t)$ is largest for LES $64^3$ due to the reasons explained in the text. }
\label{fig:epsilon_t} 
\end{figure}

In Fig.~\ref{fig:epsilon_t} we plot $\epsilon(t)$ vs.~$t$ for different simulations. Note that $\epsilon(t)$ is higher for lower resolution runs, which can be explained as follows. 
According to Eq.~(\ref{eq:dissipation}), $\epsilon(t)$ is affected by two factors: (a) the number of modes over which summation is performed, and (b) the change in the value of $\nu_r(k_c)$ with grid resolutions owing to its dependence on cutoff as $k_c^{-4/3}$. Note that, with decreasing grid resolutions, $\epsilon(t)$ decreases due to (a), but it increases due to (b).  We observe that $\epsilon(t)$ is affected more by (b) than by (a).  This is the reason for the observed increase in dissipation rate for LES with lower grid resolutions. 

\begin{figure}
\begin{center}
\includegraphics[scale=0.44]{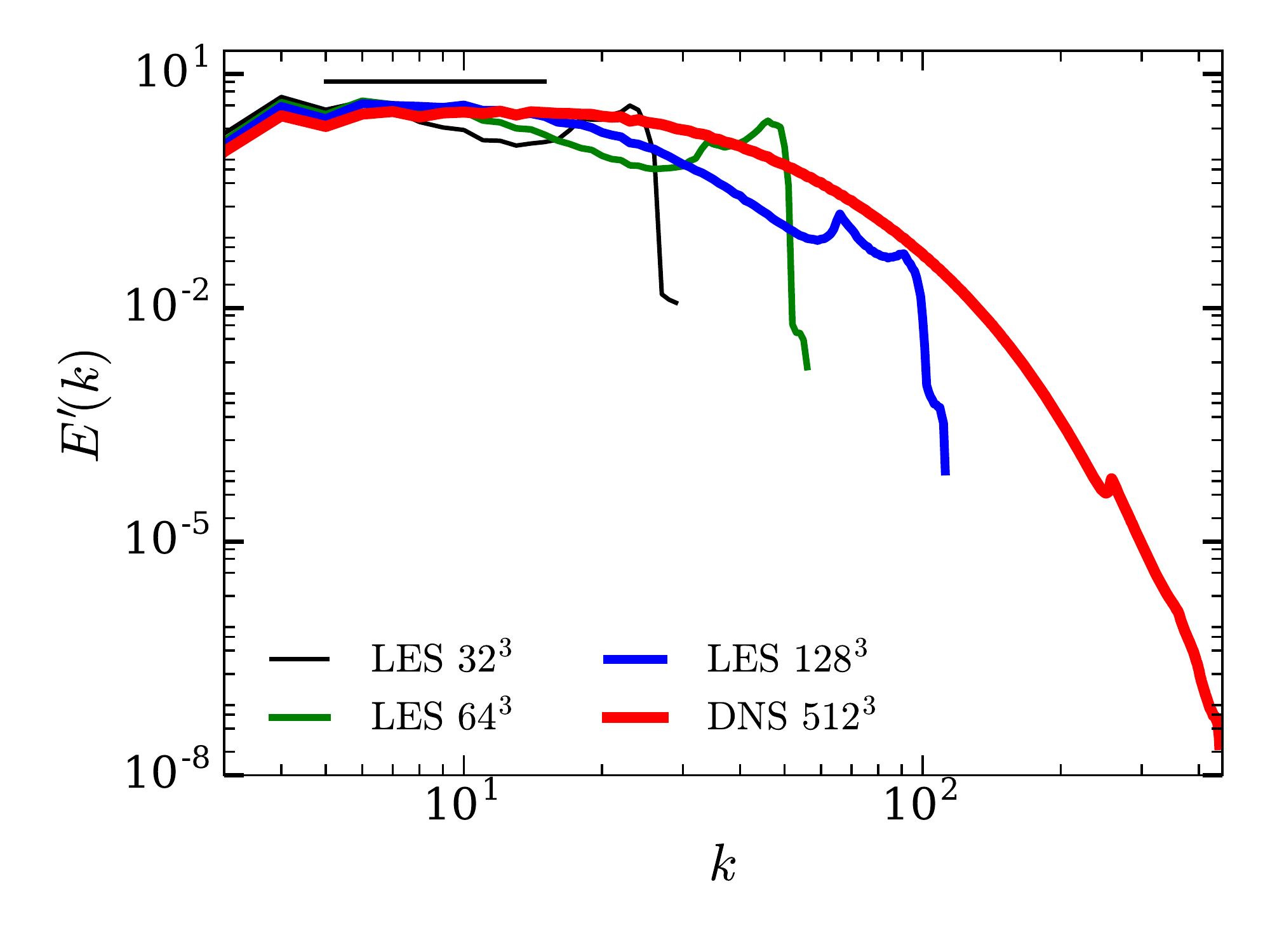}
\end{center}
\caption{ Normalized kinetic energy spectrum ${E'(k)=E_u(k)k^{5/3}\Pi^{-2/3}}$  for DNS and LES at $t=2$.  $E'(k)$ in the inertial range are approximately equal.  LES runs exhibit bumps in $E'(k)$ due to bottleneck effect.}
\label{fig:E_s} 
\end{figure}

\begin{figure}
\begin{center}
\includegraphics[scale=0.4]{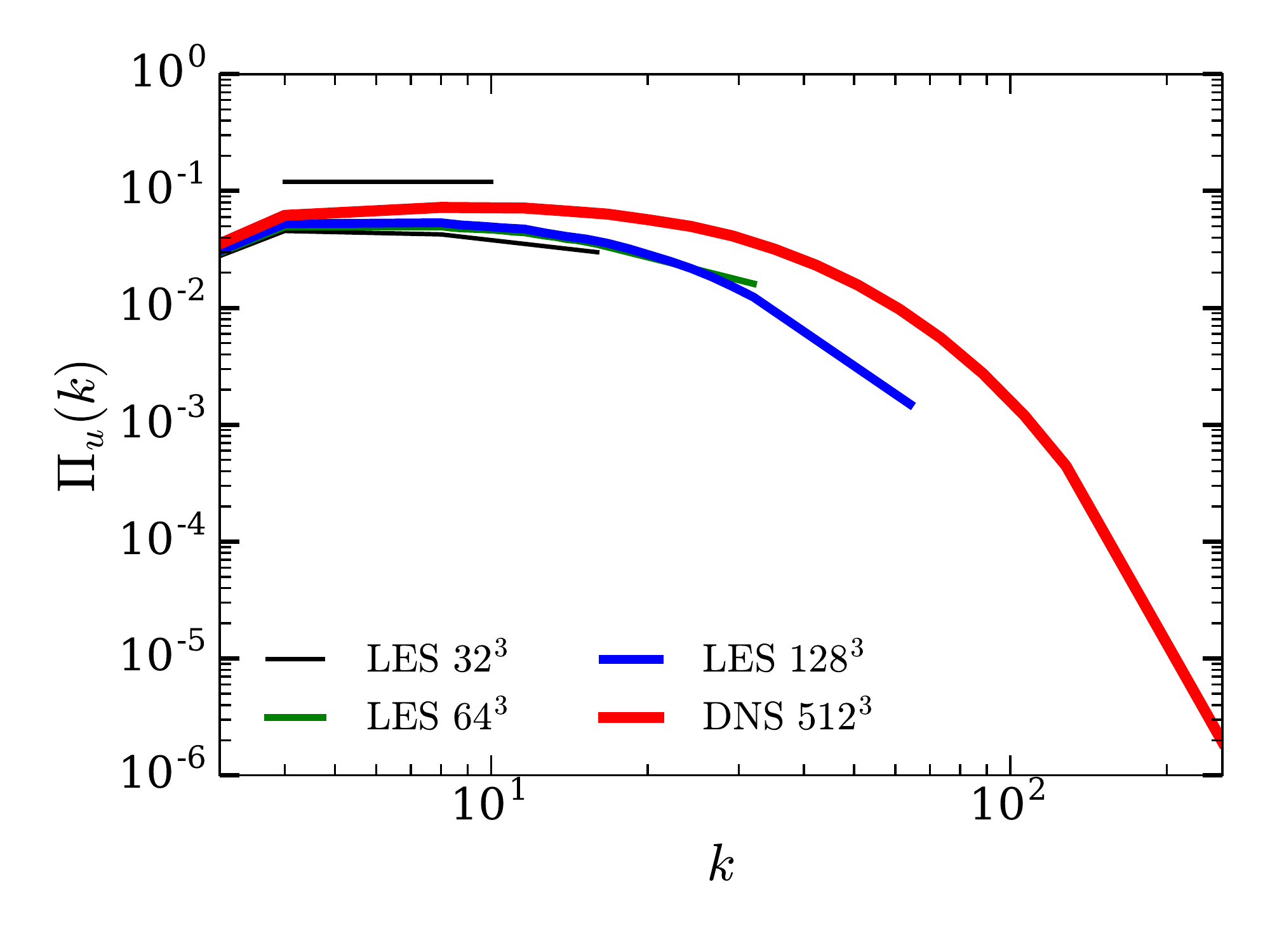}
\end{center}
\caption{Kinetic energy flux $\Pi_u(k)$ for DNS and LES at  $t=2$.  In the inertial range, $\Pi_u(k)$ are approximately equal for all the runs. }
\label{fig:Pi} 
\end{figure}

Figs.~\ref{fig:E_s} and \ref{fig:Pi} exhibit the normalized kinetic energy spectrum $E'(k) = E(k)k^{5/3}\Pi_u^{2/3}$ and the kinetic energy flux ${\Pi_u(k)}$ respectively at $t=2$.  We observe that in the inertial range, the normalised spectrum $E'(k)$ computed using DNS and LES are quite close to each other.  In addition, as shown in Fig.~\ref{fig:Pi}, the  energy flux $\Pi_u(k)$ is constant in the inertial range, consistent with the constancy of $E'(k)$.  Thus, our LES runs capture the inertial range physics quite well.  Note however that $\Pi_u(k)$ decreases as resolution of LES is lowered.  This is because the truncated large wavenumber modes eliminate some of the nonlinear triads, thus decreasing the nonlinear coupling and the energy flux.  

As discussed earlier, the dissipation rate increases as the resolution is decreased.  This is contrary to the  decrease in flux for such scenarios.   An interesting phenomenon occurs near the beginning of the dissipation rate to bring consistency among the two quantities.  To increase the energy dissipation, the system enhances $E'(k)$ near the beginning of the dissipation range as a bump as shown in Fig.~\ref{fig:E_s}.  This is the bottleneck effect, as observed earlier~\cite{Falkovich:PF1994,Verma:JPA2007}.

 \begin{figure}
\begin{center}
\includegraphics[scale=0.6]{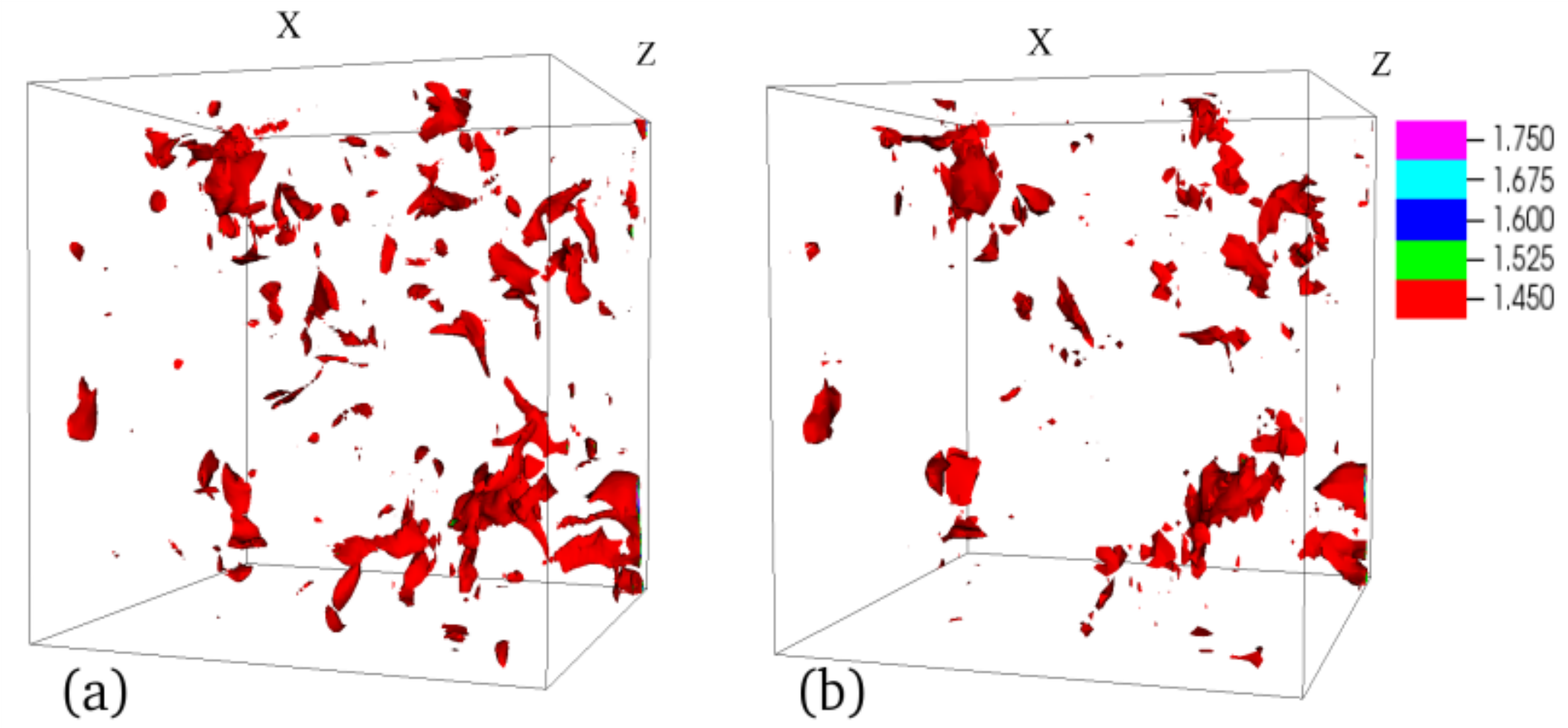}
\end{center}
\caption{ Isosurfaces of $|\mathbf{u}(\mathbf{x})|$ for (a) DNS $512^3$ and (b) LES $64^3$ at $t=2$.  Clearly, LES captures the large-scale features of DNS.}
\label{fig:VM_Iso} 
\end{figure}

After the diagnostics of spectral space, we compare the DNS and LES results in real space. For the same we present the isosurfaces of magnitude of  velocity field,  $|\mathbf{u}(\mathbf{x})|$, at a given time.  In Fig.~\ref{fig:VM_Iso} we compare the density plots of $|\mathbf{u}(\mathbf{x})|$ for DNS and and LES at $t=2$.  As is evident from the figure, the large-scale features of DNS are quite nicely captured in LES with a resolution of $64^3$.  This result, along with earlier ones, provide strong validation of our LES scheme based on renormalized viscosity.

We conclude in the next section.

\section{Conclusions}
\label{sec:Conclusion}
In the present work we have employed renormalized viscosity derived using  renormalization-group scheme for performing LES of decaying turbulence inside a periodic cubical box.  We compared the LES results with DNS results and showed that the LES with 1/8 resolution compared to DNS can capture the evolution of total energy and total dissipation rate, as well as the energy spectrum and flux.  The large-scale real-space structures are captured quite accurately by the LES. Thus, we validate our  renormalization-group based LES scheme. 

We however remark that the present renormalized viscosity is not suitable for  anisotropic and inhomogeneous turbulent flows, or in the presence of walls.  \citet{Yakhot:JSC1989} employed RG-based ideas to simulate anisotropic flows, and \citet{chasnov1991simulation} employed EDQNM stochastic model for capturing back scatter, but more sophisticated works are required for a better understanding.  

We performed our LES simulations in  a periodic box, which does not have any boundary layer.  In the boundary layers, the flow is altered due to the velocity gradients introduced by no-slip boundary condition. We need ingenious generalization of the present ideas for an employment in  the presence of walls.  One approach would be to employ kinetic energy flux locally at small enough boxes because the flux is different at different positions. We may employ the third-order structure function for such computation.  We plan to attempt such generalizations in future.
In summary, renormalization-group  based LES offers interesting set of possibilities that need to be explored in future.

\section*{ Acknowledgments} 
We thank Manohar Sharma,  Syed Fahad Anwer, and Shashwat Bhattacharya for useful discussions. The simulations were performed on the HPC system and Chaos cluster of IIT Kanpur, India. This work was supported by a research grant PLANEX/PHY/2015239 from Indian Space Research Organisation (ISRO), India. 

%\bibliographystyle{elsarticle-num}
% \bibliographystyle{elsarticle-harv}
%\bibliographystyle{elsarticle-num-names}
% \bibliographystyle{model1a-num-names}
% \bibliographystyle{model1b-num-names}
% \bibliographystyle{model1c-num-names}
% \bibliographystyle{model1-num-names}
% \bibliographystyle{model2-names}
 %\bibliographystyle{model3a-num-names}
% \bibliographystyle{model3-num-names}
%bibliographystyle{model4-names}
% \bibliographystyle{model5-names}
%\bibliographystyle{model6-num-names}

\bibliography{main.bbl}
\end{document}